\documentclass[aps,twocolumn,superscriptaddress,floatfix,nofootinbib,showpacs,amsmath,amssymb,altaffilletter]{revtex4-1}
\pdfoutput=1
\input{ds.def}
\begin{document}
\title{Results from the first use of low radioactivity argon in a dark matter search}
\date{\today}
\newcommand{\APC}{APC, Universit\'e Paris Diderot, CNRS/IN2P3, CEA/Irfu, Obs. de Paris, Sorbonne Paris Cit\'e, Paris 75205, France}
\newcommand{ \AQGSSI}{Gran Sasso Science Institute, L'Aquila AQ 67100, Italy}
\newcommand{\AQLNGS}{Laboratori Nazionali del Gran Sasso, Assergi AQ 67010, Italy}
\newcommand{\Augustana}{Department of Physics, Augustana University, Sioux Falls, SD 57197, USA}
\newcommand{\Belgorod}{Radiation Physics Laboratory, Belgorod National Research University, Belgorod 308007, Russia}
\newcommand{\Campinas}{Institute of Physics Gleb Wataghin, Universidade Estadual de Campinas, S\~ao Paulo 13083-859, Brazil}
\newcommand{\CAINFN}{Istituto Nazionale di Fisica Nucleare, Sezione di Cagliari, Cagliari 09042, Italy}
\newcommand{\CAUni}{Department of Physics, Universit\`a degli Studi, Cagliari 09042, Italy}
\newcommand{\Chicago}{Kavli Institute, Enrico Fermi Institute, and Dept. of Physics, University of Chicago, Chicago, IL 60637, USA}
\newcommand{\BHSU}{School of Natural Sciences, Black Hills State University, Spearfish, SD 57799, USA}
\newcommand{\Dubna}{Joint Institute for Nuclear Research, Dubna 141980, Russia}
\newcommand{\FNAL}{Fermi National Accelerator Laboratory, Batavia, IL 60510, USA}
\newcommand{\GEINFN}{Istituto Nazionale di Fisica Nucleare, Sezione di Genova, Genova 16146, Italy}
\newcommand{\GEUni}{Department of Physics, Universit\`a degli Studi, Genova 16146, Italy}
\newcommand{\Hawaii}{Department of Physics and Astronomy, University of Hawai'i, Honolulu, HI 96822, USA}
\newcommand{\Houston}{Department of Physics, University of Houston, Houston, TX 77204, USA}
\newcommand{\IHEP}{Institute of High Energy Physics, Beijing 100049, China}
\newcommand{\Kiev}{Institute for Nuclear Research, National Academy of Sciences of Ukraine, Kiev 03680, Ukraine}
\newcommand{\Krakow}{Smoluchowski Institute of Physics, Jagiellonian University, Krakow 30348, Poland}
\newcommand{\Kurchatov}{National Research Centre Kurchatov Institute, Moscow 123182, Russia}
\newcommand{\LLNL}{Lawrence Livermore National Laboratory, Livermore, CA 94550, USA}
\newcommand{\LPNHE}{LPNHE Paris, Universit\'e Pierre et Marie Curie, Universit\'e Paris Diderot, CNRS/IN2P3, Paris 75252, France}
\newcommand{\LSC}{Laboratorio Subterr\'aneo de Canfranc, Canfranc Estaci\'on 22880, Spain}
\newcommand{\MEPhI}{National Research Nuclear University MEPhI, Moscow 115409, Russia}
\newcommand{\MIINFN}{Istituto Nazionale di Fisica Nucleare, Sezione di Milano, Milano 20133, Italy}
\newcommand{\MIUni}{Department of Physics, Universit\`a degli Studi, Milano 20133, Italy}
\newcommand{\MSU}{Skobeltsyn Institute of Nuclear Physics, Lomonosov Moscow State University, Moscow 119991, Russia}
\newcommand{\NAINFN}{Istituto Nazionale di Fisica Nucleare, Sezione di Napoli, Napoli 80126, Italy}
\newcommand{\NAUni}{Department of Physics, Universit\`a degli Studi Federico II, Napoli 80126, Italy}
\newcommand{\PNNL}{Pacific Northwest National Laboratory, Richland, WA 99354, USA}
\newcommand{\Petersburg}{St. Petersburg Nuclear Physics Institute NRC Kurchatov Institute, Gatchina 188350, Russia}
\newcommand{\PGINFN}{Istituto Nazionale di Fisica Nucleare, Sezione di Perugia, Perugia 06123, Italy}
\newcommand{\PGUni}{Department of Chemistry, Biology and Biotechnology, Universit\`a degli Studi, Perugia 06123, Italy}
\newcommand{\Princeton}{Department of Physics, Princeton University, Princeton, NJ 08544, USA}
\newcommand{\RMTreINFN}{Istituto Nazionale di Fisica Nucleare, Sezione di Roma Tre, Roma 00146, Italy}
\newcommand{\RMTreUni}{Department of Physics and Mathematics, Universit\`a degli Studi Roma Tre, Roma 00146, Italy}
\newcommand{\USP}{Instituto de F\'{i}sica, Universidade de S\~ao Paulo, S\~ao Paulo 05508-090, Brazil}
\newcommand{\SLAC}{SLAC National Accelerator Laboratory, Menlo Park, CA 94025, USA}
\newcommand{\IPHC}{IPHC, Universit\'e de Strasbourg, CNRS/IN2P3, Strasbourg 67037, France}
\newcommand{\Temple}{Department of Physics, Temple University, Philadelphia, PA 19122, USA}
\newcommand{\Davis}{Department of Physics, University of California, Davis, CA 95616, USA}
\newcommand{\UCAS}{School of Physics, University of Chinese Academy of Sciences, Beijing 100049, China}
\newcommand{\UCLA}{Department of Physics and Astronomy, University of California, Los Angeles, CA 90095, USA}
\newcommand{\UMass}{Amherst Center for Fundamental Interactions and Dept. of Physics, University of Massachusetts, Amherst, MA 01003, USA}
\newcommand{\VTech}{Department of Physics, Virginia Tech, Blacksburg, VA 24061, USA}

\author{P.~Agnes}\affiliation{\APC}
\author{L.~Agostino}\affiliation{\LPNHE}
\author{I.~F.~M.~Albuquerque}\affiliation{\Princeton}\affiliation{\USP}
\author{T.~Alexander}\affiliation{\UMass}\affiliation{\FNAL}
\author{A.~K.~Alton}\affiliation{\Augustana}
\author{K.~Arisaka}\affiliation{\UCLA}
\author{H.~O.~Back}\affiliation{\Princeton}\affiliation{\PNNL}
\author{B.~Baldin}\affiliation{\FNAL}
\author{K.~Biery}\affiliation{\FNAL}
\author{G.~Bonfini}\affiliation{\AQLNGS}
% ...
\collaboration{The \DS\ Collaboration}\noaffiliation

\begin{abstract}
Liquid argon is a bright scintillator with potent particle identification properties, making it an attractive target for direct-detection dark matter searches. 
The \DSf\ dark matter search here reports the first WIMP search results obtained using a target of low-radioactivity argon. \DSf\ is a dark matter detector, using two-phase liquid argon time projection chamber, located at the Laboratori Nazionali del Gran Sasso. The underground argon is shown to contain \ce{^{39}Ar} at a level reduced by a factor \DSfUArArThreeNineDepletion\ relative to atmospheric argon.  We report a background-free null result from \DSfUArExposure\ of data, accumulated over  \DSfUArAfterQualityCutsLiveTime. When combined with our previous search using an atmospheric argon, the  \NinetyPerCentCL\ upper limit on the \WIMP-nucleon spin-independent cross section based on zero events found in the WIMP search regions, is  \DSfUArLimitHundredGev\ (\DSfUArLimitOneTeV, \DSfUArLimitTenTeV) for a \WIMP\ mass of \WIMPMassHundredGev\ (\WIMPMassOneTev, \WIMPMassTenTev). 

\end{abstract}
\maketitle
The existence of dark matter in the Universe is inferred from abundant astrophysical and cosmological observations~\cite{Faber:1979em, Spergel:1988fg,Clowe:2006hr}. The \DSf\ experiment searches for dark matter in the form of weakly interacting massive particles (WIMPs)~\cite{Feng:2010dz}, whose collisions with argon nuclei would produce nuclear recoils (\NRs) with tens of keV energy. Liquid argon (LAr) is a bright scintillator and allows for efficient drift and extraction of the ionization electrons. Pulse shape discrimination (\PSD) in LAr allows electron recoil (ER) events from  $\beta$-$\gamma$ backgrounds to be rejected relative to the \NR\ events expected from WIMP scattering at the \DSfAArROIEventsNumber\ level or better~\cite{Boulay:2006hu,Agnes:2015gu}.  However, atmospheric argon (\AAr) contains $\sim$1 Bq/kg of cosmic-ray produced \ce{^{39}Ar} activity~\cite{Loosli:1983bu,Benetti:2007fg}.  A source of argon with reduced \ce{^{39}Ar} activity is a crucial requirement for developing experiments that will push argon-based  WIMP dark matter direct detection searches to their highest possible sensitivity. 
This report presents the first results from a direct-detection WIMP dark matter search using a target of low radioactivity argon (\UAr), which was extracted and purified in a multi-year effort~\cite{AcostaKane:2008im, Back:2012vo, Back:2012um, Xu:2015do}. 

The \DSf\ two-phase (liquid-gas) Argon Time Projection Chamber (\LArTPC) is mounted at the center of a Liquid Scintillator Veto (\LSV) described in Ref.~\cite{Agnes:2015uo}.
The \LSV\ is instrumented with \LSVPMTsNumber\ \PMTs\ and filled with \LSVScintillatorMass\  of boron-loaded liquid scintillator.  Surrounding the \LSV\  is a \CTFWaterMass\ Water Cerenkov Veto (\WCV)  instrumented with \CTFPMTsNumber\ \PMTs. Signals from the \LSV\ and \WCV\ are used to reject events in the \LArTPC\ caused by cosmic-ray muons~\cite{Bellini:2011jd,Bellini:2012bk}, cosmogenic (muon-induced) neutrons~\cite{Bellini:2013kr,Empl:2014ih} or radiogenic neutrons and \grs\ from radioactive contamination in the detector components.

The \LArTPC\ is fully described  in Ref.~\cite{Agnes:2015gu}. A total of 38 3" \PMTs, 19 positioned at the top and  19 at the bottom of a \DSfActiveMass\ active target of \UAr\ detect primary scintillation (\SOne) and gas scintillation from drifted ionization electrons  (\STwo) resulting from ionizing radiation interactions.  The TPC drift field is \DSfDriftField\ and the extraction field is \DSfExtractionField. \PSD\ of  ER events is based on the single parameter \FNinety, the fraction of \SOne\ light detected in the first \WindowFNinety\ of the pulse. The \SOne\ and \STwo\ signals together enable 3D event localization. The transverse ($x$-$y$) position is determined from the hit pattern of the \STwo\ signal on the top PMT array, while the vertical ($z$) position is inferred from the drift time separating the \SOne\ and \STwo\ signals.
The S1 response is corrected for $z$-dependence, and the S2 response is corrected for radial dependence, normalizing both to the respective centers of the detector.  Other spatial dependencies are not significant (S1 radial dependence is $<$3\%, S2  $z$-dependence is consistent with an electron drift lifetime $>$5 ms). 
The fully corrected zero-field \TPC\ photoelectron yield with \UAr\ at the \ce{^{83m}Kr} peak energy is \DSfAArNullFieldLightYield,  2\% higher than that quoted in Ref.~\cite{Agnes:2015gu}, due to small changes in the baseline finding and pulse identification algorithms.

\begin{figure}[!t]
\includegraphics[width=\columnwidth]{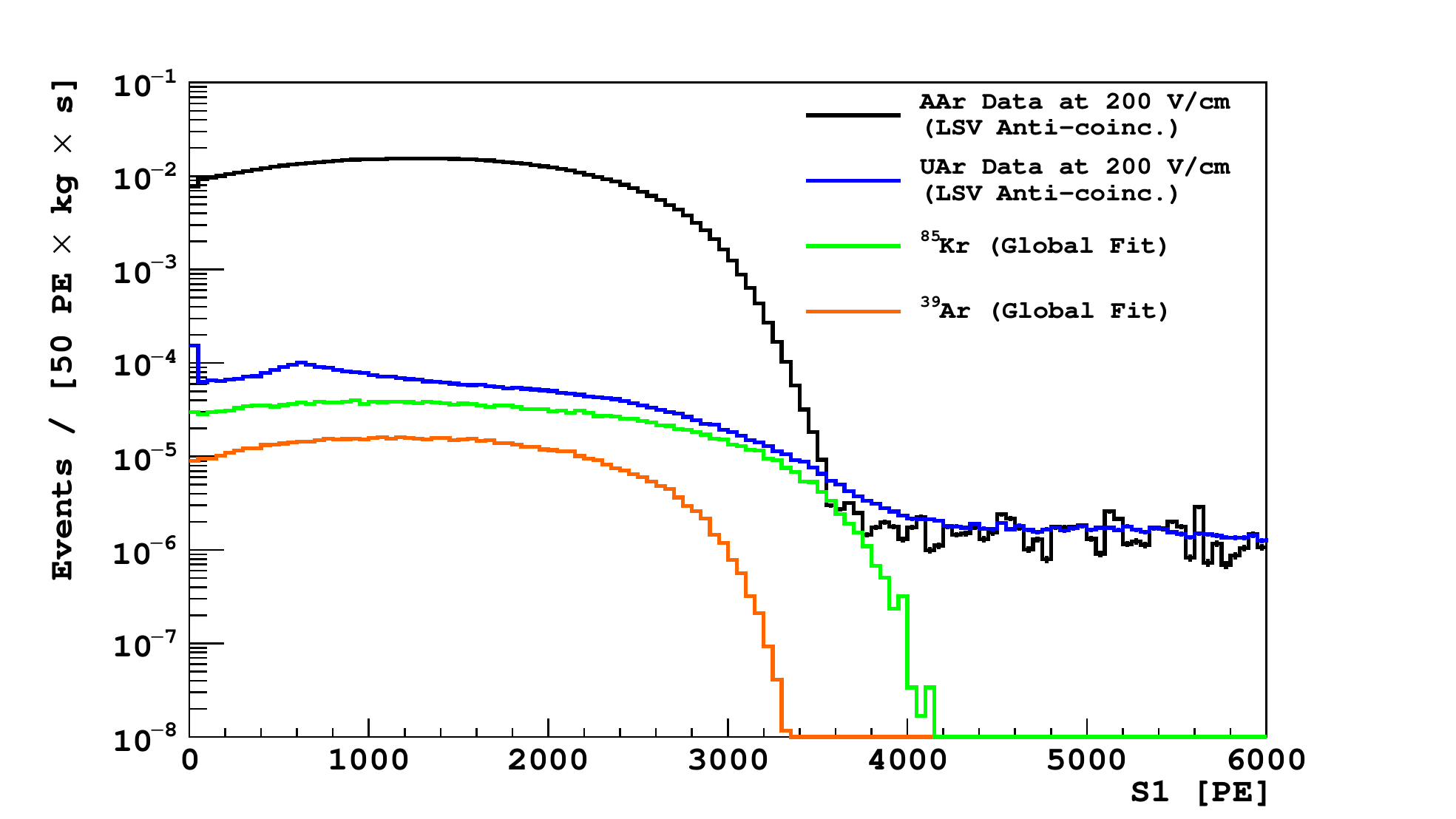}
\caption{Live-time normalized  \SOne\ pulse integral spectra from single-scatter events in \AAr\ (black) and \UAr\ (blue) taken with \DSfDriftField\ drift field.  Also shown are the \ce{^{85}Kr} (green) and \ce{^{39}Ar} (orange) levels as inferred from a \MC\ fit.  Note the peak in the lowest bin of the \UAr\ spectrum, which is due to \ce{^{37}Ar} from cosmic-ray activation.  The peak at $\sim$\SI{600}{\pe} is due to \gr\ Compton backscatters.}
\label{fig:DSf-AArUArSpectra}
\end{figure} 

Fig.~\ref{fig:DSf-AArUArSpectra} compares the \UAr\ and \AAr\ data of the \SOne\ pulse integral spectrum. A $z$-cut (residual mass of \DSfSpectraFitMass) has been applied to remove \gr\ events from the anode and cathode windows. Events identified as multiple scatters or coincident with a  prompt signal in the \LSV\ have also been removed. To compare the ER background from  \UAr\ with that from \AAr, a \Geant~\cite{Agostinelli:2003fg,Allison:2006cd} \MC\ simulation of the \DSf\ \LArTPC, \LSV, and \WCV\ detectors was developed.  The simulation accounts for material properties, optics, and readout noise, and also includes a model for \LAr\ scintillation and recombination.  The \MC\ is tuned to agree with the high statistics \ce{^{39}Ar} data taken with \AAr~\cite{Agnes:2015gu}. A simultaneous MC fit to the S1 spectrum taken with field off (see Fig.~\ref{fig:DSf-AArUArSpectra-Zero-Field} in Appendix~\ref{app}), S1 spectrum with field on, and the $z$-position distribution of events, determines the \ce{^{39}Ar} and  \ce{^{85}Kr} activities in the \UAr\ to be  \DSfUArArThreeNineActivity\ and  \DSfUArKrEightFiveActivity\ respectively. The fitted \ce{^{39}Ar} and  \ce{^{85}Kr} activities are also shown in Fig.~\ref{fig:DSf-AArUArSpectra}. The uncertainties in the fitted activities are dominated by systematic uncertainties from varying fit conditions. The \ce{^{39}Ar} activity of the \UAr\ corresponds to a reduction by a factor of \DSfUArArThreeNineDepletion\ relative to \AAr. This is significantly beyond the upper limit of 150 established in~\cite{Xu:2015do}.

An independent estimate of the  \ce{^{85}Kr} decay rate in \UAr\ is obtained by identifying $\beta$-$\gamma$ coincidences from the  \KrEightFiveExcitedDecayBR\ decay branch to metastable \ce{^{85m}Rb} with mean lifetime \DSfUArRbEightFiveMMeanLife. This method gives a decay rate of \ce{^{85}Kr} via \ce{^{85m}Rb}  of \DSfUArKrEightFiveMRateFromCoincidences\ in agreement with the value \DSfUArKrEightFiveMRateFromFit\ obtained from the known branching ratio and the spectral fit result. The presence of \ce{^{85}Kr} in \UAr\ is unexpected. We have not attempted to remove krypton from the \UAr, although cryogenic distillation would likely do this very effectively.  The \ce{^{85}Kr} in \UAr\ could come from atmospheric leaks or from natural fission underground, which produces \ce{^{85}Kr}  in deep underground water reservoirs at specific activities similar to  those of \ce{^{39}Ar}~\cite{Lehmann:2010fy}.

\begin{figure}[t!]
\centering
\includegraphics[width=\columnwidth]{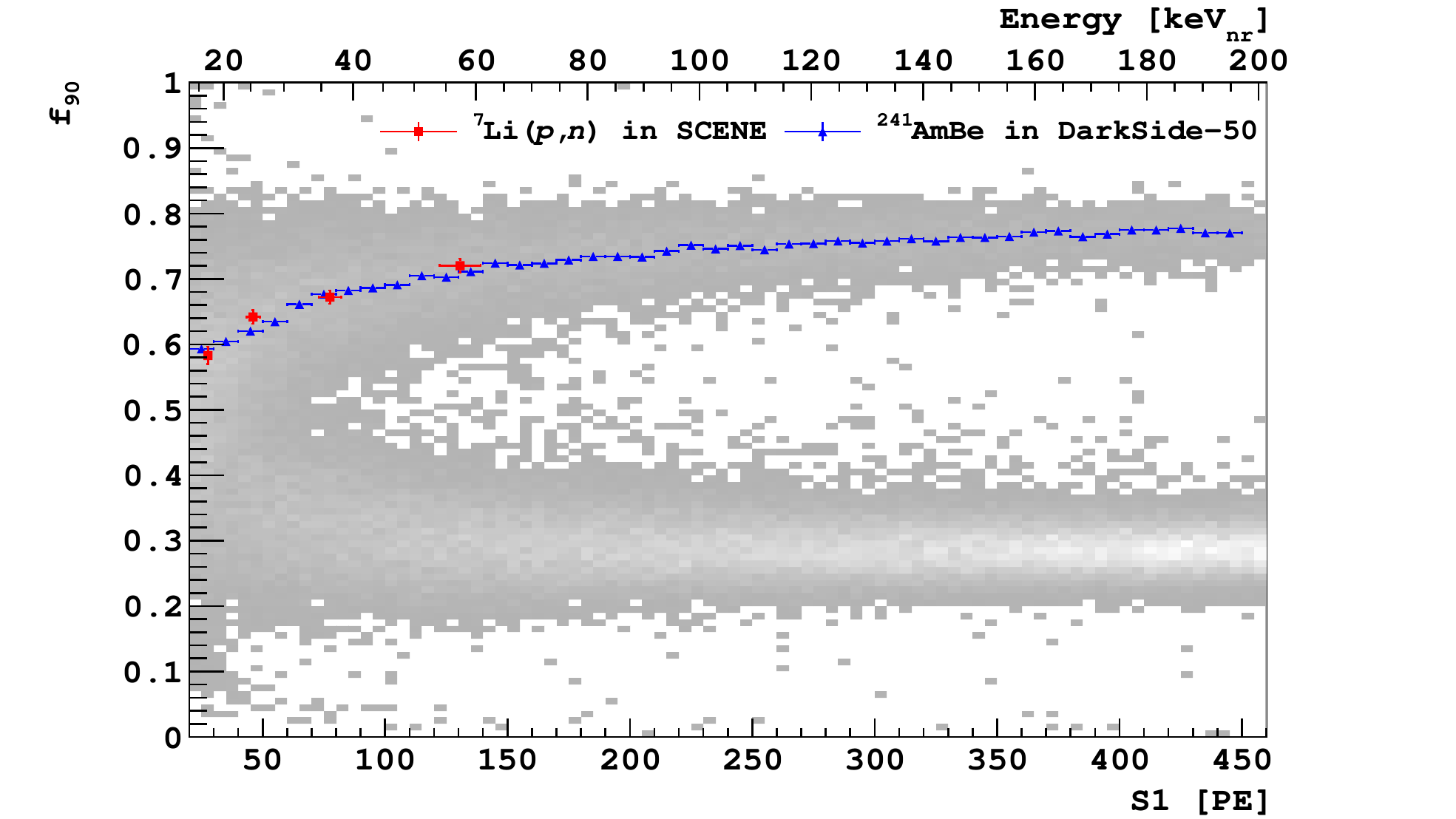}
\caption{\FNinety\ \NR\ median vs. \SOne\ from a high-rate {\it in situ} \AmBe\ calibration (blue) and scaled from \SCENE\ measurements (red points). Grey points indicate the upper NR band from the \AmBe\ calibration and lower \ER\ band from $\beta$-$\gamma$ backgrounds. Events in the region between the \NR\ and \ER\ bands are due to inelastic scattering of high energy neutrons, accidentals, and correlated neutron and \gr\ emission by the \AmBe\ source.} 
\label{fig:DSf-UArAmBeDMS}
\end{figure}

As in Ref.~\cite{Agnes:2015gu}, we determine the nuclear recoil energy scale from the \SOne\ signal using the photoelectron yield of \NRs\ relative to \ce{^{83m}Kr} measured in the \SCENE\ experiment~\cite{Alexander:2013ke,Cao:2015ks}, and the zero-field photoelectron yield for \ce{^{83m}Kr} measured in \DSf. An {\it in-situ} calibration with an \AmBe\ source was also performed, allowing a check of the  \FNinety\ medians obtained for \NRs\  in \DSf\ with those scaled from \SCENE, as shown in Fig.~\ref{fig:DSf-UArAmBeDMS}.  Contamination from inelastic or coincident electromagnetic scattering cannot easily be removed from \AmBe\ calibrations, so we still derive our \NR\ acceptance from \SCENE\ data where available.

High performance neutron vetoes are necessary to exclude \NR\ events due to radiogenic or cosmic-ray produced neutrons from the WIMP search.
In the \AAr\ exposure~\cite{Agnes:2015gu}, the vetoing efficiency of the \LSV\ was limited to 98.5$\pm$0.5\% by deadtime considerations given  the \LSVOldCFourteenRate\  of \ce{^{14}C} in the scintillator, resulting from the unintended use of trimethylborate (\TMB). For the \UAr\  dataset, the \LSV\ contains a scintillator mixture of low radioactivity \TMB\ from a different supplier at \LSVNewTMBConcentration\ concentration by mass. As a result, the \ce{^{14}C} activity in the \LSV\ scintillator is now only \LSVNewCOneFourRate. 

Neutron capture on \ce{^{10}B} in the scintillator occurs with a \LSVNewNeutronCaptureMeanLife\ lifetime through two channels ~\cite{Agnes:2015uo, Wright:2011ig}:
\begin{eqnarray*}
\ce{^{10}B} + n \rightarrow \alpha\,(\BTenNeutronCaptureGroundDecayAlphaEnergy) +\ce{^7Li}\,\,\,	& (\mbox{BR:}\,\BTenNeutronCaptureGroundDecayBR)\\
\ce{^{10}B} + n \rightarrow \alpha\,(\BTenNeutronCaptureExcitedDecayAlphaEnergy) + \ce{^7Li^*}	& (\mbox{BR:}\,\BTenNeutronCaptureExcitedDecayBR)\\
\ce{^7Li^*}& \rightarrow \ce{^7Li} + \gamma\,(\BTenNeutronCaptureExcitedDecayGammaEnergy)	
\end{eqnarray*}

The reduced radioactivity of the \LSV\ scintillator allowed us to operate with a veto window of 6 times the neutron capture lifetime and a threshold low enough to veto on the signal from the $\alpha$ and \ce{^7Li} (g.s) capture channel. Using \AmBe\ calibration data we measured that this signal is quenched to 30$\pm$5\,PE, well above our analysis threshold of 6\,PE. The \BTenNeutronCaptureExcitedDecayGammaEnergy\ \gr\ accompanying the \ce{^7Li^*} channel gives at least 240\,PE and is easily detected. From \AmBe\ data and \MC\ simulations, we estimate a detection efficiency of \LSVNewNeutronsRadiogenicRejectionEfficiencySoleCapture~\cite{Agnes:2015uo} for radiogenic neutrons  when using the neutron capture signals only. This estimate is a lower limit since the calculation neglects the neutron thermalization signal from the scintillator. The main detection inefficiency is due to the fraction of the neutron captures on \ce{^1H} in which the \HOneNeutronCaptureGammaEnergy\ de-excitation \gr\ is fully absorbed in inert materials rather than in the scintillator. 

The data for the \WIMP\ search were acquired using a simple majority trigger requiring a threshold number of channels in the \LArTPC\ to present hits within a \DSfTriggerWindow\ window.  The trigger efficiency is essentially \DSfTriggerEfficiency\ for \NRs\ in our \WIMP\ search region.  We perform a non-blind physics analysis, where the \LArTPC\ event selection and data analysis procedures are intentionally kept as similar as possible to those used in the \AAr\ exposure~\cite{Agnes:2015gu}.  After data quality cuts, we obtain \DSfUArAfterQualityCutsLiveTime\ of \WIMP\ search data with the \UAr. 

Events are further required to have only one valid and unsaturated \SOne, one valid \STwo\ pulse with position-corrected value greater than \CutMWMin, and up to one ``\SThree" pulse, due to \STwo-induced photoionization of the cathode.  A pulse is identified as \SThree\ if the time difference between \STwo\ and the pulse  matches the maximum drift time. Additionally, we remove events in which the \SOne\ light is abnormally concentrated in a single \PMT, which could be due to  an afterpulse or to a Cherenkov interaction in a PMT window piled up with a normal \SOne\ pulse.  The much lower \ce{^{39}Ar} rate in \UAr\ revealed a higher fraction of spurious events, leading us to adjust the cut to reject 5\% of events rather than 1\% as in the \AAr\ run.

The remaining events are subject to being vetoed as neutron-associated. Events are vetoed if the \LSV\ detected a prompt signal near the \LArTPC\ trigger time or if the \LSV\ detected a delayed signal above \SI{3}{PE} within \SI{200}{\us} after a \TPC\ interaction (delayed neutron captures). Events with \LSV\ activity preceding the \LArTPC\ signal by up to \pvcuttime\ are also vetoed to account for possible delayed neutron events in the TPC. Finally, all \LArTPC\ events are rejected for \mvcutdelay\ after a \TPC\ trigger in coincidence with any large-amplitude muon-like event in the \WCV\ or \LSV\, to eliminate delayed neutrons possibly produced by the muon.  

With the same $z$-cuts in the \TPC\ as in Ref.~\cite{Agnes:2015gu}, a fiducial mass of \DSfFiducialMass\ remains.  No $x$-$y$ cut is applied because the \PSD, $z$-cut and veto cuts are more than adequate to remove the $\gamma$-ray background strongly concentrated at the boundaries of the sensitive volume. Surface backgrounds from $\alpha$-emitters of the natural radioactive decay chain​s​  have been identified and studied, but none of these survive the standard cuts to give background in the \WIMP\ search region at the present background and exposure levels.

\begin{figure}[t!]
\centering
\includegraphics[width=\columnwidth]{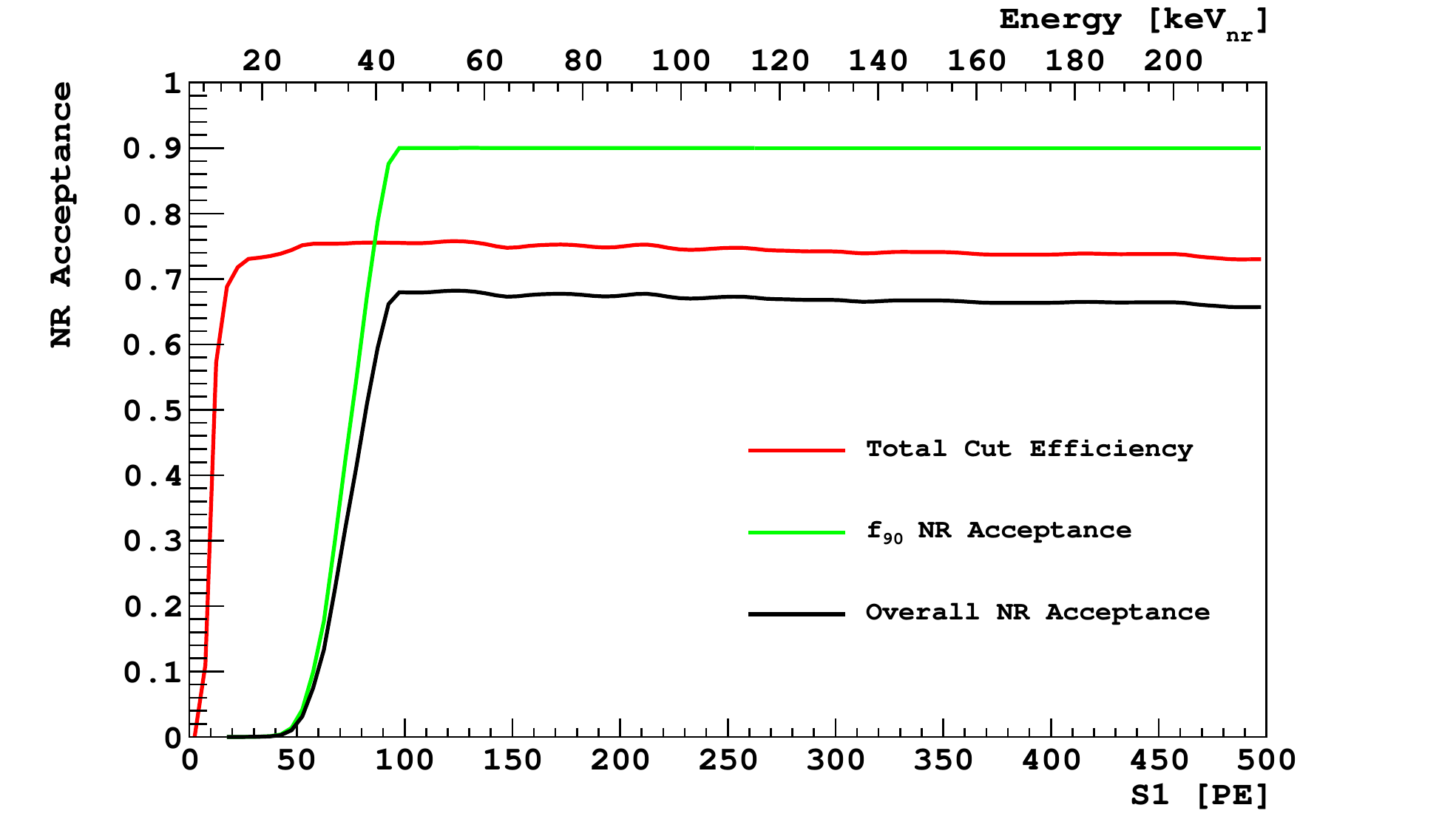}
\caption{Combined acceptance of all TPC and veto cuts (red), acceptance of the \FNinety\ \NR\ cut (green) and the final cumulative \NR\ acceptance in \UAr\ data (black).}
\label{fig:DSf-UArEfficiency}
\end{figure}

\begin{figure}[!t]
\includegraphics[width=\columnwidth]{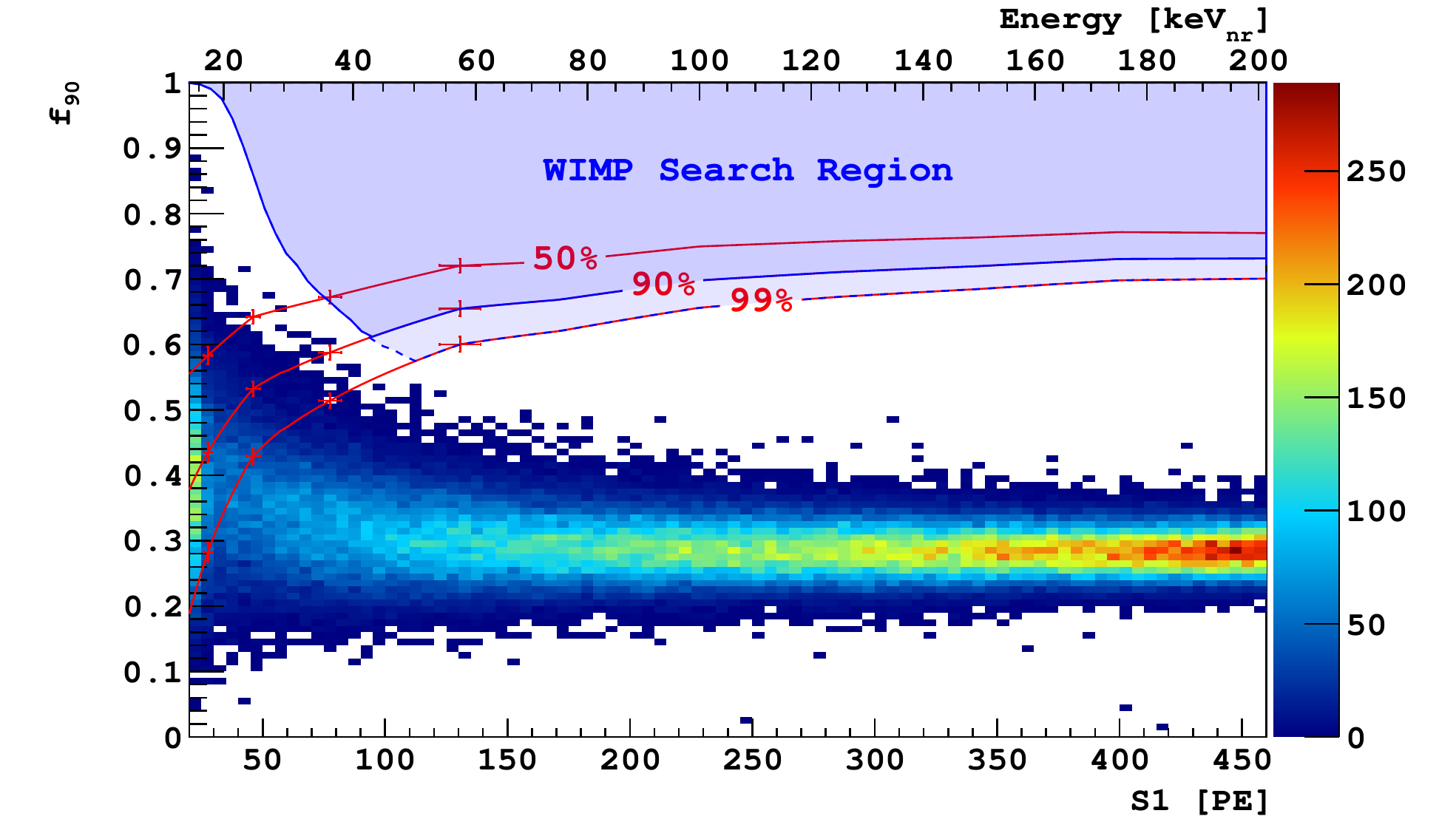} 
\caption{Distribution of events in the \FNinety\ vs \SOne\ plane surviving all cuts  in the energy region of interest.  Shaded blue with solid blue outline: WIMP search region. The red points (with their uncertainties) are derived from the \SCENE\ measurements of \NR\ acceptance. The \FNinety\ acceptance contours are drawn by connecting the red points and extending the contours using \DSf\  \AmBe\ data (see text). Lighter shaded blue with dashed blue line show that extending the WIMP search region to \DSfDMSFNinetyNineNineUpperLimit\ \FNinety\ \NR\ acceptance is still far from \ER\ backgrounds. }
\label{fig:DSf-UAr2015DMS}
\end{figure}

The combined acceptance of all TPC and veto cuts to retain single-scatter \NR\ events is shown as a function of \SOne\ in Fig.~\ref{fig:DSf-UArEfficiency}. The acceptance is \DSfUArPhysicsCutsEfficiency\ and approximately independent of \SOne\ above \CutSOMin, with the major loss being due to the dead time from the delayed neutron capture veto cut. The distribution of the \DSfUArROIEventsNumber\ events in the \FNinety\ vs.~\SOne\ plane which remain after all cuts is shown in Fig.~\ref{fig:DSf-UAr2015DMS}.  

As was done for the \AAr\ exposure, the WIMP search region is defined as a region in the  \FNinety\ vs.~\SOne\ plane having known high acceptance for \NRs\ and low expected leakage of single-scatter ER events, with an energy region of interest of \DSfUArROIPERange\ in \SOne\ (\DSfUArROIEnergyRange). NR acceptance curves are established using the median \FNinety\ values for \NRs\ measured in the \SCENE\ experiment~\cite{Alexander:2013ke,Cao:2015ks}, inserted into a statistical model for the \FNinety\ distribution, as described in Refs.~\cite{Hinkley:1969ek,Boulay:2006hu,Agnes:2015gu}.  Above \SCENEEnergyMax, where \SCENE\ data are unavailable, the \NR\ \FNinety\ medians are taken from \DSf\  \AmBe\ calibration data (see Fig.~\ref{fig:DSf-UArAmBeDMS}).  

The expected single-scatter ER leakage is calculated from the same statistical model for the \ER\ \FNinety\ distribution as described in Ref.~\cite{Agnes:2015gu}, fitted to the high statistics \ce{^{39}Ar} data from the \AAr\ exposure, and scaled to the number of events in the \UAr\ data sample.  The WIMP search region is then defined by intersecting the \DSfDMSFNinetyNineZeroUpperLimit\ \NR\ acceptance line with the curve corresponding to a leakage of less than \DSfDMSArThreeNineBackgroundCondition\ from the single-scatter \ER\ background into the \WIMP\ search region. This procedure predicts a total of less than 0.1 leakage events. As can be seen in Fig.~\ref{fig:DSf-UAr2015DMS}, there are in fact no events in the \WIMP\ search region in the present \UAr\ exposure.

We can compare the observed number of ``neutron events''---events within the WIMP search region that pass the TPC cuts and are accompanied by veto signals---with our MC prediction.   We do not observe any neutron events in the present exposure. In the previous \AAr\ exposure of \CutVPAArLiveDay\ ~\cite{Agnes:2015gu} we observed two. One of the \AAr\ neutron events was classified as cosmogenic based on its \WCV\ and \LSV\ signals. Combining the two exposures, we observe \DSfGlobalNeutronsRadiogenicObserved\ radiogenic neutron event in \DSfGlobalLiveTime\ of data, which is in agreement with our MC prediction of \DSfGlobalNeutronsRadiogenicExpected\ before the veto cuts.  MC simulations for the UAr exposure predict that \DSfGlobalNeutronsRadiogenicExpectedPassVetoCuts\ radiogenic neutrons would produce events in the TPC and remain un-vetoed.  The un-vetoed cosmogenic neutron background is expected to be small compared to the radiogenic neutron background~\cite{Empl:2014ih}.

\begin{figure}[!t]
\includegraphics[width=\columnwidth]{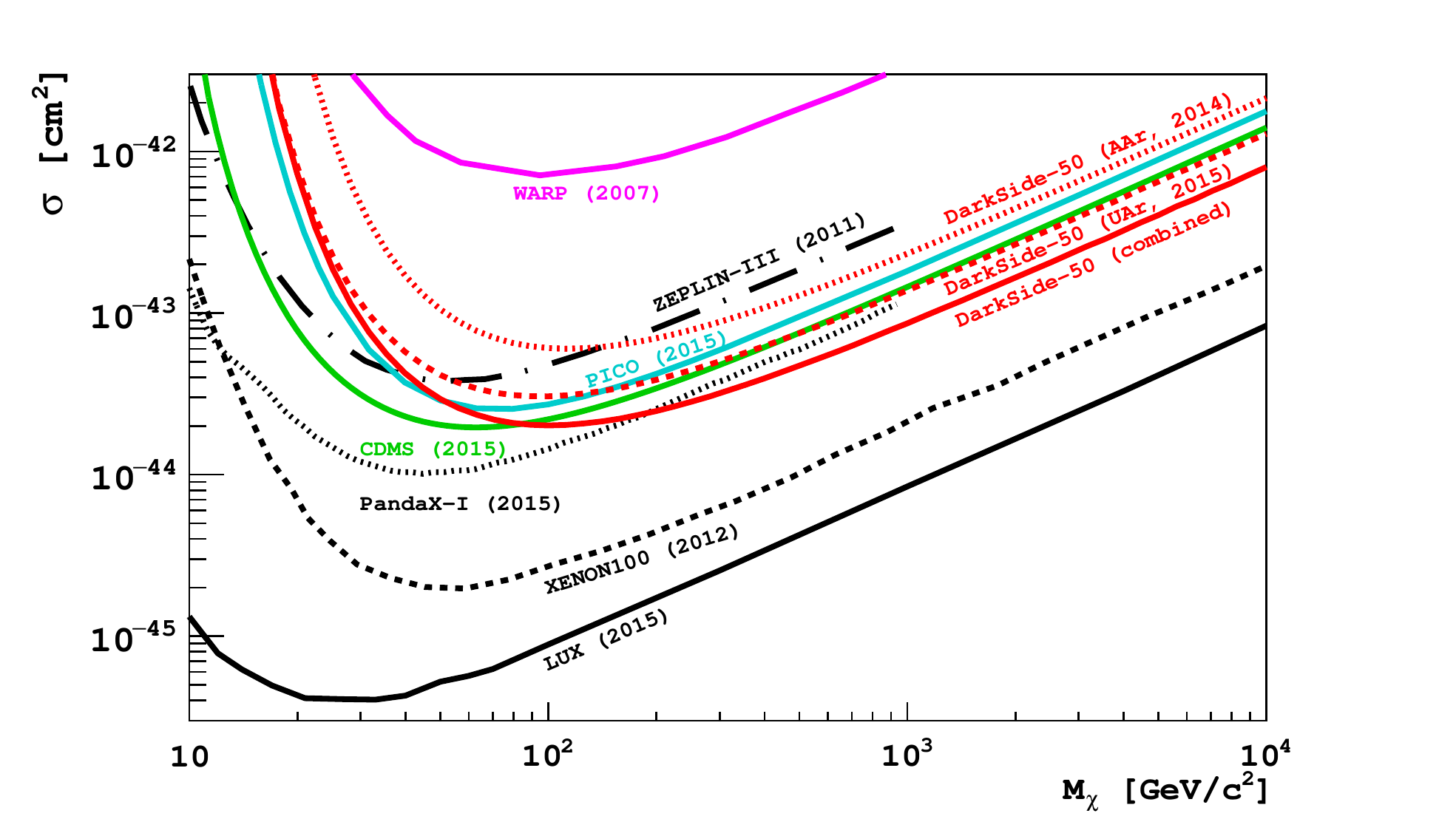}
\caption{Spin-independent WIMP-nucleon cross section \SI{90}{\percent}~C.L. exclusion plots for the \DSf\ \AAr\ (dotted red) and \UAr\ campaigns (dashed red), and combination of the \UAr\ and \AAr~\cite{Agnes:2015gu} campaigns (solid red).  Also shown are results from LUX~\cite{Akerib:2015th}(solid black), XENON100~\cite{Aprile:2012kx} (dashed black), PandaX-I~\cite{Xiao:2015ko} (dotted black), CDMS~\cite{Agnese:2015cv} (solid green), PICO~\cite{Amole:2015eu} (solid cyan), ZEPLIN-III~\cite{Akimov:2012jn} (dash dotted black) and WARP~\cite{Benetti:2008kd} (magenta).}
\label{fig:DSf-UAr2015Exclusion}
\end{figure}

Dark matter limits from the present exposure are determined from our WIMP search region using the standard isothermal galactic WIMP halo parameters (\VelocityEscapeSymbol=\VelocityEscapeValue, \VelocityNaughtSymbol=\VelocityNaughtValue, \VelocityEarthSymbol=\VelocityEarthValue, \RhoDMSymbol=\RhoDMValue, see~\cite{Agnes:2015gu} and references cited therein).  Given the background-free result shown above, we derive a \NinetyPerCentCL\ exclusion curve corresponding  to the observation of \ZeroBackgroundNinetyPerCentCLEventsLimit\ for spin-independent interactions.  The null result of the \UAr\ exposure sets the upper limit on the \WIMP-nucleon spin-independent cross section of 3.1$\times$ 10$^{-44}$ cm$^2$ (1.4$\times$ 10$^{-43}$ cm$^2$, 1.3$\times$ 10$^{-42}$ cm$^2$) for a \WIMP\ mass of \WIMPMassHundredGev\ (\WIMPMassOneTev, \WIMPMassTenTev). When combined~\cite{Yellin:2011wo} with the  null result of our previous \AAr\ exposure, we obtain an upper limit of \DSfUArLimitHundredGev\ (\DSfUArLimitOneTeV, \DSfUArLimitTenTeV) for a \WIMP\ mass of \WIMPMassHundredGev\ (\WIMPMassOneTev, \WIMPMassTenTev).  Fig.~\ref{fig:DSf-UAr2015Exclusion} compares these limits to those obtained by other experiments.

The \DSf\ detector is currently accumulating exposure in a stable, low-background configuration with the characteristics described above.  We plan to conduct a \DSfPlannedRunTime\ dark matter search with increased calibration statistics and several improvements in data analysis (see Fig.~\ref{fig:DSf-UAr2015DMS-StwoCut} in Appendix~\ref{app}). These first results show that \UAr\ can significantly extend the potential of argon for \WIMP\ dark matter searches. The \ER\ rejection previously demonstrated in \AAr\ data and the reduction of  \ce{^{39}Ar}  shown here already imply that \UAr\ exposures of at least 5.5 tonne-yr can be made free of  \ce{^{39}Ar} background.  

\begin{acknowledgments}
The DarkSide-50 Collaboration would like to thank LNGS and its staff for invaluable  technical and logistical  support.  This report is based upon work supported by the US NSF (Grants \grant{PHY}{0919363}, \grant{PHY}{1004072}, \grant{PHY}{1004054}, \grant{PHY}{1242585}, \grant{PHY}{1314483}, \grant{PHY}{1314507} and associated collaborative grants; Grants \grant{PHY}{1211308} and \grant{PHY}{1455351}), the Italian Istituto Nazionale di Fisica Nucleare, the US DOE (Contract Nos.~\grant{DE}{FG02-91ER40671} and \grant{DE}{AC02-07CH11359}), and the Polish NCN (Grant \grant{UMO}{2012/05/E/ST2/02333}). We thank the staff of the Fermilab
Particle Physics, Scientific and Core Computing Divisions for their support.  We acknowledge the financial support from the UnivEarthS Labex program of Sorbonne Paris Cit\'e (\grant{ANR}{10-LABX-0023} and \grant{ANR}{11-IDEX-0005-02}) and from the S\~ao Paulo Research Foundation (FAPESP).
\end{acknowledgments}

\appendix
\section{Field off spectra and \STwo/\SOne\ cut}
\label{app}

\begin{figure}[h!]
\includegraphics[width=\columnwidth]{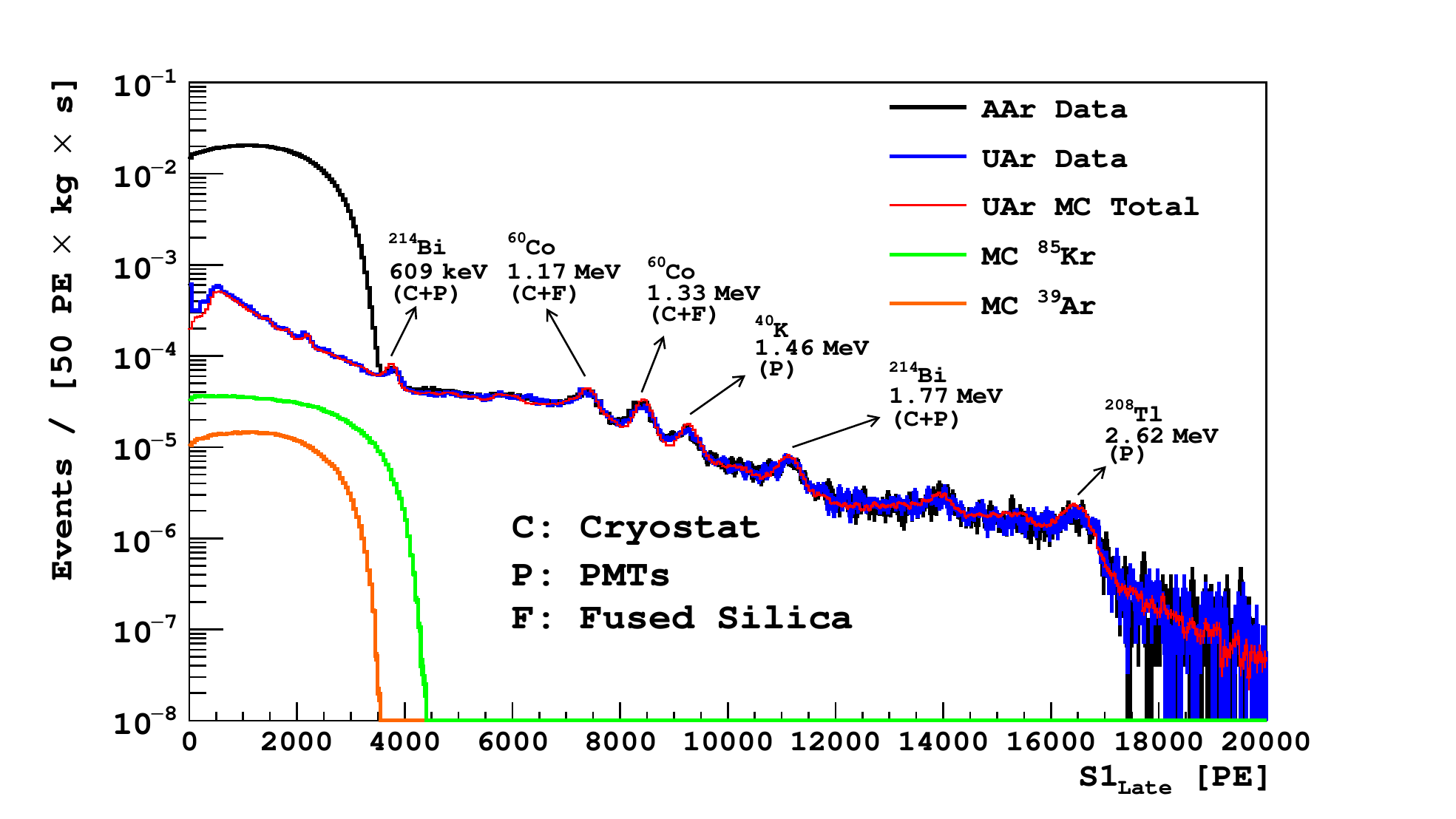} 
\caption{Comparison of the measured field off spectra for the \UAr\ (blue) and \AAr\ (black) targets, normalized to exposure.  Also shown are the  MC fit to the \UAr\ data (red) and individual components of \ce{^{85}Kr} (green) and \ce{^{39}Ar} (orange) extracted from the fit. }
\label{fig:DSf-AArUArSpectra-Zero-Field}
\end{figure} 

Fig.~\ref{fig:DSf-AArUArSpectra-Zero-Field} compares the measured field off spectra for the \UAr\ (blue) and \AAr\ (black) targets, normalized to exposure.  The horizontal axis (``\SOne-late") is the integral of the \SOne\ pulse from \WindowFNinety\ to 7$\mu$s, which includes $\sim$70\% of the total S1 light for electron recoils (\ERs). Despite the sacrifice of photoelectron statistics, use of \SOne-late avoids distortion of the spectra by digitizer saturation at high \SOne\ values (\SOne$>$\SI{2E3}{\pe}) and, with the asymmetry correction for \SOne\ described above, gives a net improvement in the pulse height resolution.  The background \gr\ lines originate from identified levels of \ce{^{238}U}, \ce{^{232}Th}, \ce{^{40}K}, and \ce{^{60}Co} in the detector construction materials and are consistent with the expectations from our materials screening.  The repeatability in the positions of the peaks in the  \AAr\ and \UAr\ data shows the stability of the detector system as a whole. 

\begin{figure}[!t]
\includegraphics[width=\columnwidth]{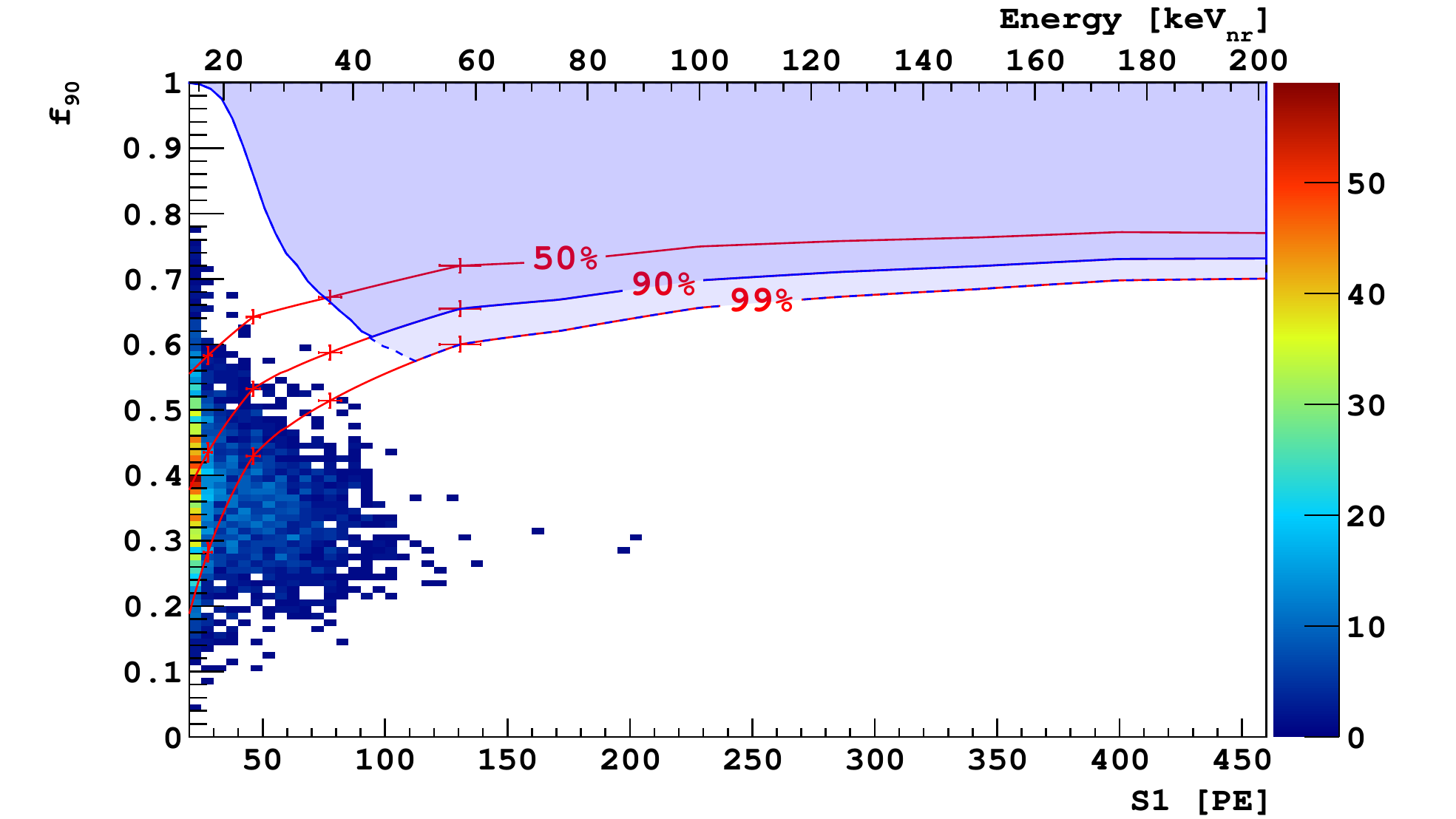} 
\caption{ Distribution of events in the \FNinety\ vs \SOne\ plane which survive all quality and physics cuts (including veto cuts), and which in addition survive a radial cut and a \STwo/\SOne\ cut.  Shaded blue with solid blue outline: WIMP search region.  Lighter shaded blue with dashed blue line show that extending the WIMP search region to \DSfDMSFNinetyNineNineUpperLimit\ \FNinety\ \NR\ acceptance is still far away from \ER\ backgrounds..}
\label{fig:DSf-UAr2015DMS-StwoCut}
\end{figure}

Fig.~\ref{fig:DSf-UAr2015DMS-StwoCut} demonstrates available improvements in background rejection, which we do not utilize in this analysis.  When adding  an \STwoSoneRatio\ cut (requiring that \STwoSoneRatio\ be lower than the median value for \NRs) and also xy fiducialization (requiring the reconstructed radius to be less than \CutRSMax), we obtain an even greater separation between the events surviving the selection and the previously defined WIMP search region.  Should a signal appear in the region of interest, the \STwoSoneRatio\ parameter would provide a powerful additional handle in understanding its origin.

\end{document}